\documentclass[trackchanges]{aastex701}

\usepackage{academicons}
\usepackage{orcidlink}
\usepackage{hyperref}
\usepackage{amsmath}
\usepackage{amssymb}
\usepackage{amsfonts}
\usepackage{graphicx}
\usepackage{bm}
\usepackage{subfigure}
\usepackage{subcaption}
\usepackage[inline]{enumitem}

\newcommand{\update}[1]{\textcolor{black}{#1}}

\begin{document}

\title{On the Acceleration of Pulsar Timing computations using Normalising Flows and Parallelisation}

\author{Churchil Dwivedi \orcidlink{0000-0002-8804-650X}}
\affiliation{Astronomy and Astrophysics Division, Physical Research Laboratory, Thaltej Campus, Thaltej, Ahmedabad 380059, Gujarat, India}
\email[show]{churchil@prl.res.in}  

\author{Hiya Shah \orcidlink{0009-0008-6426-6672}}
\affiliation{School of Engineering and Applied Science, Ahmedabad University, Ahmedabad 380009, Gujarat, India}
\email[]{hiya.s5@ahduni.edu.in} 

\author{Hemanga Tahbildar \orcidlink{0009-0002-1036-9306}}
\affiliation{Department of Physics, IISER Bhopal, Bhauri Bypass Road, Bhopal, 462066, Madhya Pradesh, India}
\email[]{hemanga1998@gmail.com} 

\begin{abstract}
Single-Pulsar Noise Analysis (SPNA) and Gravitational Wave (GW) searches done on Pulsar Timing Array (PTA) datasets have everlastingly suffered from the computational bottleneck arising due to high dimensionality and multi-modality of the PTA likelihood landscape, along with strong correlations amongst various single-pulsar noises and ensemble-level common noise processes. We addressed this outstanding issue by employing a Normalising Flow-based Preconditioned Monte-Carlo sampling technique implemented in the \texttt{POCOMC} package, for the first time on PTA-specific computations, and comparing the achieved acceleration with the widely used \texttt{PTMCMCSAMPLER} and \texttt{DYNESTY} packages. We further investigated the acceleration achieved via parallelisation over an increasing array of communicating nodes on a high-performance computing (HPC) resource, by employing the \texttt{PARALLEL\_BILBY} architecture with \texttt{DYNESTY}. We tested the acceleration on realistic long baseline simulated datasets with SPNA and Common Red Noise (CRN) analysis. We found \texttt{PARALLEL\_BILBY} to be the most efficient in parallelisation, achieving a runtime of $\sim$$10\,\rm min$ and $\sim$$100\,\rm min$ with 16 nodes for spatially uncorrelated and Hellings and Downs-correlated CRN searches, respectively. \texttt{POCOMC} outperforms in single node performance requiring only $\sim$$10\,\rm h$ for correlated search. \texttt{PTMCMCSAMPLER} was found to be the least efficient. We envisage \texttt{POCOMC} to be of great importance for PTA analyses, without requiring any GPU or HPC support, while also performing ensemble-level GW searches within a manageable time span. These results have everlasting implications with increasing data volumes and need to incorporate more complicated models, which were otherwise beyond reach due to the associated computational costs.

\end{abstract}


\section{Introduction}\label{sec:Introduction}
\setcounter{footnote}{0}
\setcounter{figure}{0}
The pursuit of scientific excellence is one full of bumpy landscapes and unmarked lanes with the hope of an impending confluence. As mankind's understanding of the natural phenomena increases, it discovers places that were otherwise latent. However, such discoveries often arrive with an implicit complication of developing sophisticated experiments for measuring previously unknown effects and dealing with large data sets to characterise them appropriately. Pulsar Timing Arrays (PTAs: \citet{Detweiler1979, FosterBacker1990}) are one such experiments that leverage the high rotational stability of millisecond pulsars (MSPs) via precise monitoring of their electromagnetic emissions. This technique, called pulsar timing \citep{Hobbs+2006}, involves tracking the rotation of MSPs with high accuracy by measuring the times of arrival (ToAs) of their pulsed emissions, and correcting for various astrophysical effects \citep{Edwards+2006}. This can be mathematically expressed as
\begin{equation}\label{eq:intro-1}
    t_{\rm{obs}} \simeq t_{\rm{em}} + \sum_n\Delta_n(t_{\rm em}) + \widetilde{\mathcal{N}},
\end{equation}
where $t_{\rm em}$ is the emission time in the reference frame of the pulsar, while $t_{\rm obs}$ is the observed time. The summation encapsulates the propagation delays that the photons encounter in their path from the pulsar to the Earth, such as delays due to the binary motion of the pulsar, finite light travel time, dispersion and scattering through the \update{ionised Interstellar Medium (ISM)} and Interplanetary Medium (IPM), orbital motion of the Sun and Earth, and clock-corrections relating observatory time standards to the Solar System Barycenter (SSB) reference \citep{Hobbs+2006}. A prescription of these delays forms the timing model ephemeris of a pulsar. The term $\widetilde{\mathcal{N}}$ refers to stochastic noise processes that can either be white, such as radiometer noise, or red, such as noise due to the turbulent \update{ISM} or inherent stochasticity associated with pulsar's rotation (referred to as the spin noise). These stochastic processes are popularly modeled via Fourier-domain Gaussian Processes (GPs) in the reduced-rank approximation \citep{vanHaasteren+Vallisneri2014}, and the corresponding delays introduced to ToAs can be written as
\begin{equation}\label{eq:intro-2}
    \Delta_{\rm GP} = \left(\frac{\nu_{\rm ref}}{\nu}\right)^\chi\sum_{i=1}^{N_c}a_i\cos{[2\pi f_i(t_i-t_{\rm ref})]} + b_i\sin{[2\pi f_i(t_i-t_{\rm ref})]},
\end{equation}
where $\chi$ is the chromatic index ($\chi=0$ for spin noise, $\chi=2$ for DM\footnote{The integrated free-electron column density along the line of sight, given by $\rm{DM}=\int_\ell n_e \mathrm d\ell$} noise), $\nu$ is the observing frequency, $f_i=i/\rm{T_{span}}$ is the Fourier-frequency for a dataset with time baseline $\rm{T_{\rm span}}$, $t_{\rm ref}$ is a reference epoch, and $N_c$ is sufficiently large to effectively capture the noise characteristics of a pulsar. PTAs aim to accurately characterise these noise processes via Single-Pulsar Noise Analysis (SPNA) techniques, wherein they are treated as perturbations to the original pre-noise timing solution obtained using packages such as \texttt{TEMPO2} \citep{Edwards+2006} or \texttt{PINT} \citep{Luo+2021, Susobhanan+2024}. In this \textit{linear} regime, we can write the timing residuals for a pulsar as
\begin{equation}\label{eq:intro-3}
    \bm{\mathrm r} \simeq \bm{\mathrm{M}\epsilon}+\bm{\mathrm Fa},
\end{equation}
where $\bm{\mathrm r}$ is a vector of timing residuals with size $N_{\rm ToA}$, $\bm{\mathrm M}$ represents the $(N_{\rm ToA}\times N_{\rm TM})$ pulsar design matrix containing partial derivatives of $\bm{\mathrm r}$ relative to $N_{\rm TM}$ timing model parameters, $\bm{\epsilon}$ is a vector of timing model deviations with size $N_{\rm TM}$, $\bm{\mathrm a}$ \update{represents a vector of Fourier coefficients of stochastic noise processes with size $2kN_c$, and $\bm{\mathrm F}$ is an $(N_{\rm ToA}\times 2kN_c)$ matrix of the noise basis vectors.} Here $k$ is the number of red noise processes included in the noise model of a pulsar. \update{Deterministic signals can also be included in this setting, but we omit their discussion here without any loss of generality (see the discussion in \citet{Nobleson+2026, GundersonCornish2025})}. We can write the likelihood function of timing residuals in this framework \citep{Lentati+2014} as
\begin{equation}\label{eq:intro-4}
    \ln\bm{\mathcal{L}}=-\frac{1}{2}\bm{\mathrm r}^T\bm{\mathrm N}^{-1}\bm{\mathrm r}-\frac{1}{2}\ln{\det(2\pi\bm{\mathrm N})},
\end{equation}
with
\begin{equation}\label{eq:intro-5}
    N_{ij}=\mathcal{F}^2(\sigma_{ij}^2+\mathcal{Q}^2)\,\delta_{ij} + \mathcal{E}^2\mathcal{U}_{ij},
\end{equation}
where $\mathcal{F}$, $\mathcal{Q}$ and $\mathcal{E}$ represent the \texttt{EFAC}, \texttt{EQUAD} and \texttt{ECORR} white noise parameters, respectively, $\sigma$ is the ToA uncertainty, $i,\,j$ are the ToA indices, and $\bm{\mathcal{U}}$ is a \update{sparse block-diagonal matrix with $1$ for ToAs of the same epoch and $0$ otherwise. While \texttt{EFAC} and \texttt{EQUAD} are used to scale the ToA uncertainties to account for radiometer noise, \texttt{ECORR} is used to account for pulse-jitter \citep{Kulkarni+2024, Kikunaga+2024}, which manifests as a correlated process across sub-banded ToAs of the same epoch but is uncorrelated across different epochs. Hence, it is modeled using the sparse block-diagonal implementation.} These white noise parameters are generally used in groups as per the receiver-backend configurations \citep{Srivastava+2023, Nobleson+2026}. In general, $N_{\rm ToA},N_c\gg1$, such that fitting for all the red noise coefficients is extremely compute intensive. Therefore, we consider $\bm{a}\sim\mathcal{N}(0, \bm{\Phi})$, such that we can obtain the analytically marginalised likelihood \citep{vanHaasteren+Levin2013} as
\begin{equation}\label{eq:intro-6}
    \ln\bm{\bm{\Lambda}}=-\frac{1}{2}\bm{\mathrm r}^T\bm{\mathrm C}^{-1}\bm{\mathrm r}-\frac{1}{2}\ln{\det(2\pi\bm{\mathrm C})},
\end{equation}
where
\begin{equation}\label{eq:intro-7}
    \bm{\mathrm C} = \bm{\mathrm N} + \bm{\mathrm F\Phi}\bm{\mathrm F}^T.
\end{equation}
The inverse in Eq.\eqref{eq:intro-6} can be efficiently computed using \update{the Woodbury identity \citep{Lentati+2013, vanHaasteren+Vallisneri2014}} as
\begin{equation}\label{eq:intro-8a}
    \bm{\mathrm C}^{-1} = \bm{\mathrm N}^{-1} - \bm{\mathrm N}^{-1}\bm{\mathrm F}\left(\bm{\mathrm \Phi}^{-1} + \bm{\mathrm F}^T\bm{\mathrm N}^{-1}\bm{\mathrm F}\right)\bm{\mathrm F}^T\bm{\mathrm N}^{-1},
\end{equation}
and the determinant can be computed as
\begin{equation}\label{eq:intro-8b}
    \det\bm{\mathrm C} = \frac{\det\bm{\mathrm N}\times \det\bm{\mathrm \Phi}}{\det\left(\bm{\mathrm \Phi}^{-1} + \bm{\mathrm F}^T\bm{\mathrm N}^{-1}\bm{\mathrm F}\right)}.
\end{equation}

Traditionally, these GPs are modeled with a Fourier-domain power-law kernel, such that we can write
\begin{equation}\label{eq:intro-9}
    \Phi_{kl\alpha\beta} = \mathcal{S}(f_k, \bm{\theta}) \, \delta_{kl} \, \Gamma_{\alpha\beta} \, /\, \rm{T_{\rm span}},
\end{equation}
where $\bm{\theta}\in\{A,\gamma\}$ is the set of hyper-parameters for GPs, $k,\,l=1,2,3,\dots,N_c$, represent the Fourier basis indices, $\alpha,\,\beta=1,2,3,\dots,N_{\rm psr}$, represent the pulsar indices for a PTA with $N_{\rm psr}$ pulsars, and $\mathcal{S}$ is the power-spectral density (PSD) of the process, given by
\begin{equation}\label{eq:intro-10}
    \mathcal{S}(f, A,\gamma)=\frac{A^2}{12\pi^2}\left(\frac{f}{f_{\rm yr}}\right)^{-\gamma}\,\rm{yr}^3,
\end{equation}
where $f_{\rm yr}=1\,\rm{yr}^{-1}$ is a reference frequency. The quantity $\Gamma_{\alpha\beta}$ is called the Overlap Reduction Function (ORF) of the timing residuals. In order to search for a stochastic Gravitational-Wave Background (GWB), traditional PTA analyses look for Common Uncorrelated Red Noise (CURN), which represents a time-correlated and spatially-uncorrelated noise process common to the ensemble of pulsars, such that $\bm{\mathrm C}$ is a block-diagonal matrix with individual blocks corresponding to a red noise process with common hyper-parameters $\{\log_{10}A_{\rm CURN},\gamma_{\rm CURN}\}$, but independent Fourier coefficients across pulsars. \update{Such a process indicates the presence of a time-correlated noise process common to the PTA ensemble, and is commonly used as a reference model against which the strength of inter-pulsar correlations can be assessed \citep{Chen+2021, Antoniadis+2022}.} \update{For an isotropic, unpolarised stochastic GWB with the standard tensor polarisations, originating due to the incoherent superposition of GW emissions from supermassive black-hole binaries (SMBHBs) in their leading-order radiation reaction-dominated inspiraling phase, the ORF is given by the characteristic Hellings and Downs (HD) function \citep{HellingsDowns1983},}
\begin{equation}\label{eq:intro-11}
    \Gamma_{\alpha\beta} = \frac{1}{2} - \frac{1}{4}\left(\frac{1-\zeta_{\alpha\beta}}{2}\right) + \frac{3}{2}\left(\frac{1-\zeta_{\alpha\beta}}{2}\right) \ln{\left(\frac{1-\zeta_{\alpha\beta}}{2}\right)}+\frac{1}{2}\delta_{\alpha\beta},
\end{equation}
where $\zeta_{\alpha\beta}$ is the sky-projected angular separation between a pulsar pair $(\alpha,\beta)$ in the array. \update{In particular, the PSD of ToA residuals due to such a GWB process for a population of circular SMBHBs follows a power-law behavior with $\gamma_{\rm GWB}=13/3$ \citep{Chen+2021}}. Therefore, the search of a GWB signal in PTA datasets requires spatial correlations to be computed for each pulsar pair in the ensemble, along with its spectral characterisation. The presence of these correlations is thus considered the detection proxy for a GWB \citep{Agazie+2023, Antoniadis+2023c, Reardon+2023, Xu+2023, Miles+2025b}. \\

\update{We can now write the Bayes' Theorem as}
\begin{equation}\label{eq:intro-12}
    \bm{\mathcal{P}}(\bm{\theta}\,|\,d)=\frac{\bm{\bm{\Lambda}}(d\,|\,\bm{\theta})\,\bm{\Pi}(\bm{\theta})}{\mathcal{Z}(d)},
\end{equation}
where $\bm{\mathcal{P}}$ is the posterior distribution of hyper-parameters, $\bm{\Pi}$ is the prior and $\mathcal{Z}$ is the Bayesian evidence, defined as the integral of the likelihood over the prior volume. The evidence values are used to compare two models using Bayes' Factors \citep{Jeffreys1961}, given by
\begin{equation}\label{eq:intro-13}
    \ln{\mathcal{B}}_{12} = \ln{\mathcal{Z}_1} - \ln{\mathcal{Z}_2}.
\end{equation}

The PTA likelihood is traditionally implemented using the \texttt{ENTERPRISE} package \citep{Ellis+2020, Johnson+2024}. \update{In order to estimate the Bayesian evidence, Nested Sampling algorithms \citep{Skilling2004} using packages such as \texttt{DYNESTY} \citep{Speagle2020} have recently been employed, particularly due to the acceleration provided by them in performing expensive PTA likelihood calculations.} However, these algorithms provide only an approximation of the posterior density, therefore, they're restricted to the model selection process. The hyper-parameter posteriors are estimated using Markov Chain Monte-Carlo techniques \citep{Metropolis+1953, Hastings1970, Earl+2005, Sharma2017} with suitable thinning to avoid parameter auto-correlations. These techniques are implemented in various flavors, such as the parallel-tempering approach in \texttt{PTMCMCSAMPLER} package \citep{Ellis+2017} and the affine-invariant approach in \texttt{EMCEE} package \citep{Foreman-Mackey+2013}. These packages are vastly used in PTA literature for SPNA and GWB searches \citep[e.g.][]{Chalumeau+2022, Agazie+2024}. \\

As evident from Eq.\eqref{eq:intro-7}, the covariance matrix is, in general, high-dimensional ($N_{\rm ToA}\gg1$) and dense (as $N_c\gg1$ for red noises, and with increased complexity in the presence of \texttt{ECORR}), making the calculations substantially compute intensive, even with the optimisations from Eqs.\eqref{eq:intro-8a}-\eqref{eq:intro-8b}. This necessitates the use of advanced techniques to overcome this computational challenge, especially with increasing data volumes and pulsar array sizes with time. In this work, we compare the traditional Bayesian computation method using \texttt{PTMCMCSAMPLER}, with the \texttt{PARALLEL\_BILBY} architecture \citep{Smith+2020} employed in recent PTA literature \citep[e.g.][]{Miles+2025b}, and a relatively new technique of Preconditioned Monte-Carlo sampling implemented in the \texttt{POCOMC} package \citep{Karamanis+2022a, Karamanis+2022b}. We aim to study the relative differences in computational times with these methods and benchmark the \texttt{POCOMC} technique, since its use in the PTA community is rather limited. This is \update{particularly important} for IPTA analyses \citep{Verbiest+2016} where computation times limit the complexities in GWB model architectures, even for SPNA when multiple PTA datasets are combined \citep{Agazie+2024}.\\

The rest of the paper is organised as follows. In Section \ref{sec:Techniques}, we give a brief overview of the computational techniques being used in this work. In Section \ref{sec:Analysis}, we provide details of the analysis framework adopted to compare the techniques. We present the results of our study in Section \ref{sec:Results} and discuss their implications, followed by a summary in Sections \ref{sec:Discussion} and \ref{sec:Summary}, respectively.

\section{Brief Overview of the techniques}\label{sec:Techniques}
In this section, we give a brief overview of the architectures behind the Bayesian analysis techniques used in this work. We defer the reader to the relevant literature whenever further details are warranted. For each of the techniques, we also highlight the potential of parallelisability using high-performance computing (HPC) architectures to substantiate the discussion in the later sections.

\subsection{\texttt{PTMCMCSAMPLER}}\label{subsec:PTMCMC}
The \texttt{PTMCMCSAMPLER} package\footnote{\url{https://github.com/nanograv/PTMCMCSampler.git}} is based on the Parallel Tempering (PT) Monte-Carlo technique, which is particularly suitable for sampling complex multi-modal posteriors, where the conventional Metropolis-Hastings sampling \citep{Metropolis+1953, Hastings1970} method falls short (for a detailed review, refer to \citet{Speagle2020}). In particular, PT involves running multiple MCMC chains with an artificial \textit{temperature} ladder in parallel to create a sequence $\{\mathcal{P}_T\}$ of target distributions, such that
\begin{equation}\label{eq:ptmcmc-1}
    \mathcal{P}_{T_i}(\bm{\theta}) \propto \Pi_{T_i}(\bm{\theta})\,\mathcal{L}^{1/{T_i}}(\bm{\theta})\,\forall \, i\in[1, N_{\rm T}],
\end{equation}
where $N_{\rm T}$ refers to the number of temperature ladders. It can be assessed that $\mathcal{P}\to\Pi$ as $T\to\infty$. Therefore, hotter chains can explore the flatter parts and colder ones can explore the peaks in the complex likelihood landscape, with periodic temperature swapping to ensure detailed balance. For more details on PT, refer to \citet{Earl+2005}. \update{\texttt{PTMCMCSAMPLER} is uniquely powerful in combining PT with a dynamically blended mixture of different proposal distributions\footnote{\update{In MCMC sampling, a proposal distribution gives the probability of the next step in the Markov chain, based on the current state of the chain.}}. In particular, it allows for three types of proposals -- Adaptive Metropolis (AM: to dynamically update the steps based on parameter correlations), Single-Component Adaptive Metropolis (SCAM: to fine-tune individual parameters to maintain a stable acceptance globally), and Differential Evolution (DE: to effectively navigate through multiple local extrema in the likelihood landscape), wherein their relative weights are dynamically adjusted to traverse the desired target geometry efficiently.} Additionally, to achieve parallelisation across communicating CPUs, \texttt{PTMCMCSAMPLER} natively supports communication via Message Passing Interface (MPI: \citet{MPI}). We note that several recent advancements in PT implementations, such as Accelerated PT with Neural Transports \citep{Zhang+2025}, Automatic PT \citep{Jin+2024} and High-Throughput Multiple PT \citep{Hosseini+2018} also exist in the literature, but their use in pulsar timing is rather scarce, and are hence not discussed in this work.\\

In this study, we restrict ourselves to the PT treatment implemented in \texttt{PTMCMCSAMPLER} specifically due to the reasons highlighted in Section \ref{sec:Introduction}. \update{Additionally, we have considered the default initial setting of $50\%$, $30\%$ and $15\%$ weighted pool\footnote{\update{The percentages here are expressed with respect to the overall pool size, i.e., a DE weight of $50\%$ actually corresponds relative to a pool of $50+30+15=95$. Therefore, it effectively corresponds to $\frac{50}{95}\times 100\approx52.63\%$ for DE. Same holds for AM and SCAM as well.}} for DE, SCAM and AM, respectively, and let the sampler adjust them dynamically as it traverses the parameter space\footnote{\update{\texttt{PTMCMCSAMPLER} also provides the flexibility of using empirical distributions as proposals that can potentially lead to more efficient sampling for specific problems. However, we only consider the discussed proposals in this work, without any loss of generality.}}. This choice also ensures that we do not bias the sampling efficiency by fine-tuning the proposals, which can significantly hamper the MCMC acceptance rates \citep{Speagle2020}). Furthermore, we have considered a total of $10^6$ samples with a burn factor of $30\%$ based on chain inspection, to estimate the parameter posteriors. The Gelman-Rubin statistic \citep{GelmanRubin1992, BrooksGelman1998} is employed to further assess the convergence of MCMC chains. For the analyses presented in this work, we found $10^6$ samples to be enough to achieve convergence, with $\hat{R}\approx1.01$ for all cases \citep{Vehtari+2021}.}

\subsection{\texttt{PARALLEL\_BILBY}}\label{subsec:PBILBY}
The \texttt{PARALLEL\_BILBY} framework\footnote{\url{https://lscsoft.docs.ligo.org/parallel_bilby/}} presented in \citet{Smith+2020} is the extension of the original \texttt{BILBY} package\footnote{\url{http://bilby-dev.github.io/bilby/index.html}} \citep{Ashton+2019, AshtonTalbot2021, Talbot+2025} with massive parallelisation capabilities achieved via MPI through a \textit{head-worker} architecture. Its implementation is primarily done for the Bayesian inference of compact binary coalescence events in interferometric data. However, with the developments of \citet{Samajdar+2022}, an \texttt{ENTERPRISE}-friendly implementation of \texttt{PARALLEL\_BILBY} is publicly available\footnote{\url{https://github.com/anuradhaSamajdar/parallel_nested_sampling_pta.git}}, which can be used for GW analyses with PTA datasets \citep[e.g.][]{Miles+2025b}. We use this implementation for the subsequent analysis in this work, and refer to it as the \texttt{PBILBY\_PTA} method. It is primarily based on the \texttt{DYNESTY} package to calculate the Bayesian evidence and approximate posterior density via Nested Sampling. \update{In particular, \texttt{DYNESTY} involves drawing a certain number of \textit{live points} across the prior volume, estimating the likelihood of each live point and sorting them in the decreasing likelihood order. The lowest likelihood point is removed from the active pool and added to a list of \textit{dead points}. Subsequently, the prior bounds are appropriately shrunk within a boundary based on the discarded points and a new live point is proposed within this bounded region (called the \textit{prior mass}) with a higher likelihood than the last discarded live point. At any iteration $n$, the evidence is computed using the prior mass, $X_n$, as} 
\begin{equation}\label{eq:pbilby-1}
    \mathcal{Z}_n=\int_0^1\mathcal{L}(X_n)\,\mathrm dX_n,
\end{equation}
\update{and the convergence is decided by a tolerance limit that defines the effective contribution to the evidence achieved by a further shrink in the prior mass, i.e., $\Delta\mathcal{Z}_n^{n+1}=\mathcal{Z}_{n+1}-\mathcal{Z}_{n}$ as $\mathrm dX_n\to\mathrm dX_{n+1}$. In this work, we have considered $500$ live points, a tolerance of $\Delta\log\mathcal{Z}=0.1$ as the stopping criterion, and a $100$-step random walk-based live point proposal scheme \citep{Skilling2006} with a multi-ellipsoidal bounding framework \citep{Feroz+2009}.} Having said so, we'd like to highlight that the inherent parallelisation framework of \texttt{PARALLEL\_BILBY} is fairly generic and can be extended to any sampler. This is further discussed at the outset of Section \ref{subsec:POCOMC}.

\subsection{\texttt{POCOMC}}\label{subsec:POCOMC}
The Preconditioned Monte-Carlo technique implemented in the \texttt{POCOMC} package\footnote{\url{https://github.com/minaskar/pocomc.git}} relies on using Normalising Flows (NFs) to transform the geometry \citep{Papamakarios+2019} of a complicated target distribution to a simpler one (usually a Normal distribution), and then performing Sequential Monte-Carlo sampling (SMC: \citet{DelMoral+2006}) in the transformed parameter space to sample the posterior densities \citep{Karamanis+2022a}. The posterior density in the target space of $\varphi$ is related to the transformed density in a fiducial $u$-space as
\begin{equation}\label{eq:pocomc-1}
    p_{\varphi}(\varphi) = p_u\left(f^{-1}(\varphi)\right)\left|\det{\left(\frac{\partial f^{-1}}{\partial\varphi}\right)}\right|,
\end{equation}
where $\varphi=f(u)$ and $f$ is an NF-based bijective mapping between the $\varphi$-space and the $u$-space, usually parameterised by neural networks in a way such that the Jacobian in Eq.\eqref{eq:pocomc-1} is tractable. \update{The \texttt{POCOMC} package allows for two classes of such bijective mappings -- Neural Spline Flow (NSF) and Masked Auto-regressive Flow (MAF), to model complex probability distributions. MAFs generally use affine transformations, which can only modify the parameter space along straight lines in a single layer. For highly complicated distributions, this requires stacking a large number of MAF layers for effective parameter decoupling. On the other hand, NSFs divide the entire parameter space into a grid of bins and fit a smooth piecewise rational quadratic spline in each bin \citep{Durkan+2019}. This allows a single NSF layer to be significantly more capable in modeling complex topologies than an MAF layer. Since \texttt{POCOMC} primarily uses NF as a \textit{preconditioner}, i.e., a mechanism to decouple a highly correlated parameter space for effective transformation of a complex likelihood topology to as close a multivariate Normal distribution as possible, we expect NSFs to be more efficient with lesser number of layers required to achieve the desired decorrelation compared to MAFs \citep{Coccaro+2024}. Since the PTA likelihood landscape is inherently multi-modal with high parameter correlations (see Section \ref{sec:Introduction}), we have adopted the NSF-based implementation of NF in \texttt{POCOMC} throughout this work. It should be noted that while NSF inherently supports a coupling-based layering structure, the default setting in \texttt{POCOMC} allows for a masked auto-regressive layer structure, which is also adopted in the present work.} Furthermore, since \texttt{POCOMC} is built on an SMC-based sampling method, it gives an exact estimate of the Bayesian evidence along with the posterior density. This is very useful in cutting down the two-fold model selection and parameter estimation procedures adopted in current PTA studies \citep[e.g.][]{Chalumeau+2022, Nobleson+2026}. The \texttt{POCOMC} package also natively supports an MPI-based parallelisation architecture. \\

\update{The convergence criteria in \texttt{POCOMC} is decided by the temperature annealing parameter, $\beta_{\rm T}$, of the Preconditioned SMC \citep{Karamanis+2022a}. At the start, the initial distribution is essentially the prior distribution, such that $\beta_{\rm T}=0$. At each iteration, the increment in $\beta_{\rm T}$ is done via a dynamic particle allocation scheme that adaptively scales the Effective Sample Size (ESS) based on the complexity of the parameter landscape. At each step, however, the ESS is fixed to $95\%$ of the active particles, and $\beta_{\rm T}$ is incremented such that the ESS is maintained. When $\beta_{\rm T}=1$, the particles have effectively reached the target distribution, thereby signaling the convergence of the SMC run. This is followed by a \textit{post-convergence sampling} step wherein additional equal-weighted samples are drawn from the $\beta_{\rm T}=1$ state using the trained NF till the targeted sample size (represented by $n_{\rm total}$) is reached, which is then used to estimate the posterior density. The process terminates once the desired $n_{\rm total}$ is achieved. In the present work, we have considered the default setting of $n_{\rm total}=4096$.} \\

As mentioned at the outset of Section \ref{subsec:PBILBY}, since the parallelisation framework of \texttt{PBILBY\_PTA} is fairly generic, we have implemented the \texttt{POCOMC} sampler within the \texttt{PBILBY\_PTA} architecture. This is done to see whether we get any improvements upon using the \texttt{PBILBY\_PTA} acceleration compared to what we natively get with \texttt{POCOMC}. We will refer to this as the \texttt{PBILBY\_POCOMC} implementation in the subsequent discussion.

\section{Analysis Framework}\label{sec:Analysis}
In this section, we describe the framework adopted to compare the computational efficiency of the previously described techniques. We divide the analysis into two parts -- SPNA and Common Red Noise (CRN) analysis. All the analysis is done on simulated datasets, details of which are provided in the subsequent text.

\begin{figure}[!ht]
    \centering
    \includegraphics[trim=0cm 0cm 1cm 0cm, clip, width=\linewidth]{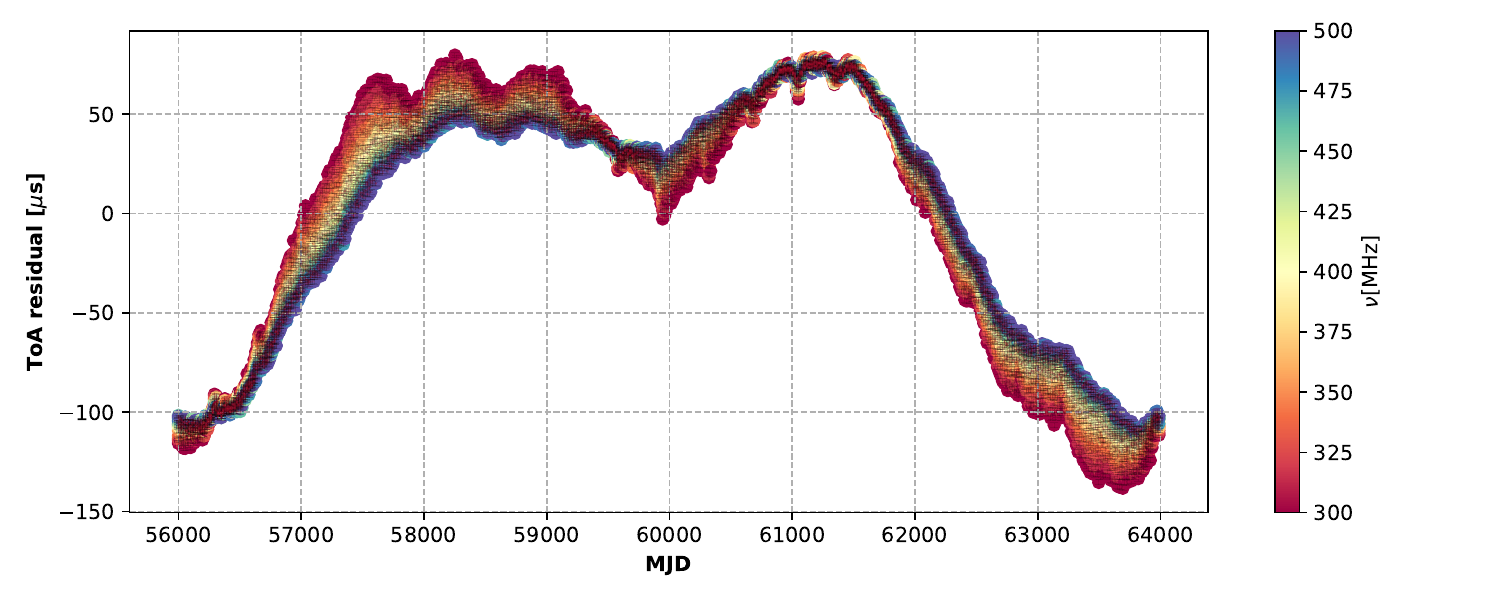}
    \caption{The simulated ToA residuals with 16 sub-bands in the $300-500\,\rm{MHz}$ observing frequency range. The ToAs have a sensitivity of $1\mu\rm s$ and $\rm T_{span}\sim20\,\rm yr$. The injected single-pulsar noise processes are as described in Section \ref{subsec:Analysis-spna}.}
    \label{fig:SPNA-ToA-residuals}
\end{figure}

\subsection{Single-Pulsar Noise Analysis}\label{subsec:Analysis-spna}
We used the \texttt{PINT} package to simulate ToAs in $300-500\,\rm{MHz}$ observing frequency range with $1\mu\rm{s}$ precision, 16 sub-bands per epoch and $\rm T_{span}\sim20\,\rm yr$ with a cadence of $14\,\rm{d}$. We used the EPTA \texttt{DR2Full+} \citep{Antoniadis+2023a} ephemeris as the initial (pre-noise) timing solution. We injected the spin, DM and solar-wind\footnote{Solar wind is a stream of charged particles that alters the free electron content along the line of sight, and hence acts as a dispersive process. It can also have a stochastic component, which can be modeled as a red noise process via GPs. For more details, refer to \citet{Tiburzi+2019, Susarla+2024}.} red noise processes with hyper-parameters $\log_{10}A=\{-13.0,\,-13.5,\,-6.8\}$ and $\gamma=\{3.5,\,3.0,\,2.5\}$, respectively. We included 16 Fourier bins below $1/\rm T_{span}$ such that the lowest bin corresponds to $1/2^{16}\rm T_{span}$. We selected $N_c=1000$ for the injected noises, such that the highest bin corresponds to $1000/\rm T_{span}$. Additionally, we incorporated an \texttt{EFAC} of 1.2, an \texttt{EQUAD} of $0.25\mu\rm{s}$ and an \texttt{ECORR} of $0.5\mu\rm{s}$ for white noise. We chose all white and red noise components so as to increase the computational complexity of the underlying noise model and put the techniques to a rather stringent test. The simulated ToA residuals are shown in Figure \ref{fig:SPNA-ToA-residuals}, where the colorbar indicates the observing frequency. \\

The \texttt{ECORR} model is implemented in two forms -- the \texttt{EcorrKernelNoise} class in \texttt{enterprise.signals.white\_signals} module which models it as a \textit{white signal}\footnote{The white signal  class in \texttt{ENTERPRISE} represents signals whose effect is only on the white-noise covariance matrix, i.e., $\bm{\mathrm{N}}$ as per Eq.\eqref{eq:intro-7}}, and the \texttt{EcorrBasisModel} class in the \texttt{enterprise.signals.gp\_signals} module which models it as a GP. Both these implementations are expected to model the same underlying phenomenon discussed in Section \ref{sec:Introduction}. \update{The former implements \texttt{ECORR} using Eq.\eqref{eq:intro-5}, thereby producing a sparse block-diagonal $\bm{\mathrm{N}}$ matrix, while the latter incorporates it's effect via the noise basis matrix $\bm{\mathrm F}$, with appropriate weights in $\bm{\mathrm{\Phi}}$ based on a given prior distribution.} The \texttt{EcorrKernelNoise} model can be computed using either the \texttt{sherman-morrison} method \citep{ShermanMorrison1949, ShermanMorrison1950} or the \texttt{fast-sherman-morrison} method \citep{vanHaasteren2023}. We use all three implementations of \texttt{ECORR} model in this work to model this process in the simulated ToAs. The red noises are recovered with $N_c=100$ bins. We performed SPNA for each of the three models using the three techniques described in Section \ref{sec:Techniques} to estimate their relative computation times. \update{Appropriate uniform priors are used for all the hyper-parameters and convergence criteria for various samplers are as described in Section \ref{sec:Techniques}.}

\subsection{Common Red Noise Analysis}\label{subsec:Analysis-crn}
For CRN analysis, we used the InPTA DR2 dataset baseline, sensitivity \citep{Rana+2025} and noise budget estimates \citep{Nobleson+2026} to create a realistic simulated dataset with an extended baseline of $\rm T_{span}\sim15\,yr$ but the same pulsar ensemble, based on the methodology presented in \citet{Pol+2021}. Additionally, we injected a GWB process with $\log_{10}A_{\rm GWB}=-14.0$ and $\gamma_{\rm GWB}=13/3$, after subtracting a realisation of the process as per the original dataset baseline from individual pulsars \citep{Tahbildar+2026}. Conventional analyses for searching common processes in the PTA ensemble reduce the dimensionality of the problem by fixing the white noise parameters to the median or maximum \textit{aposteriori} values obtained at the SPNA stage. This reduces the computational load significantly\footnote{In the InPTA DR2 dataset \citep{Rana+2025}, a total of 7 different receiver-backend configurations are present. This leads to a total of $7\times3=21$ white noises parameters for the longest baseline pulsars \citep{Nobleson+2026}. If $M$ such pulsars are present, then overall $21M$ dimensions are reduced by fixing these white noise parameters.}. We use the same methodology and search for CURN and GWB (with HD correlations) processes using each of the techniques described in Section \ref{sec:Techniques}. Additionally, we also include the \texttt{PBILBY\_POCOMC} method for these searches, as they are significantly more compute intensive than SPNA. \update{Similar to SPNA, we have used uniform priors for all the hyper-parameters and convergence criteria for various samplers are as described in Section \ref{sec:Techniques}.}

\section{Results}\label{sec:Results}
In this section, we present the results obtained for the SPNA and CRN analysis done on the simulated datasets and models described in the previous section. We use \texttt{PTMCMCSAMPLER}, \texttt{POCOMC} and \texttt{PBILBY\_PTA} architectures for SPNA, while we also include \texttt{PBILBY\_POCOMC} specifically for the CRN analysis.

\subsection{Single-Pulsar Noise Analysis}\label{subsec:Results-spna}
As the \texttt{ECORR} model is implemented in three flavors, we performed SPNA using \texttt{POCOMC} and \texttt{PBILBY\_PTA} with each of the three \texttt{ECORR} implementations. For \texttt{POCOMC}, we fixed $\rm n_{active}=256$, which corresponds to $\rm n_{particles}=256$ as per \citet{Karamanis+2022a}. For Nested Sampling inside \texttt{PBILBY\_PTA} and MCMC sampling with \texttt{PTMCMCSAMPLER}, we have adopted the same settings for the samplers and the convergence criteria for each run as described in Section \ref{sec:Techniques}. \\

\begin{figure}[!ht]
\centering
    \subfigure[\texttt{POCOMC}]{\includegraphics[trim=0cm 0cm 0cm 0cm, clip, width=0.495\linewidth]{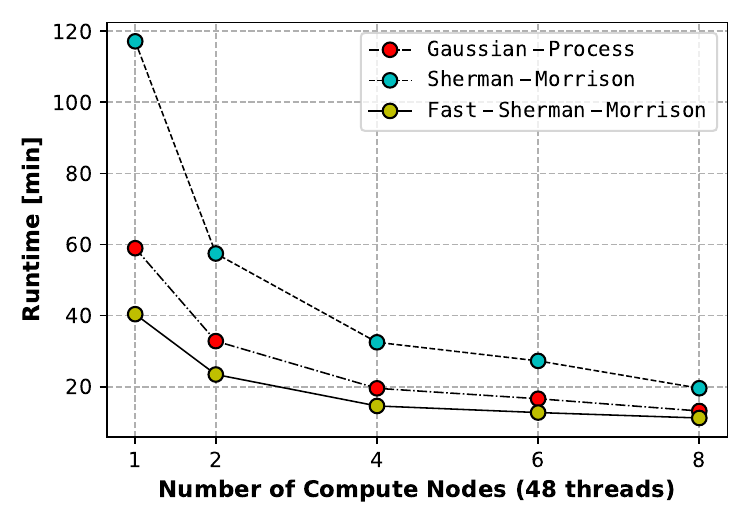}\label{fig:SPNA-ecorr-comp-1}}
    \subfigure[\texttt{PBILBY\_PTA}]{\includegraphics[trim=0cm 0cm 0cm 0cm, clip, width=0.495\linewidth]{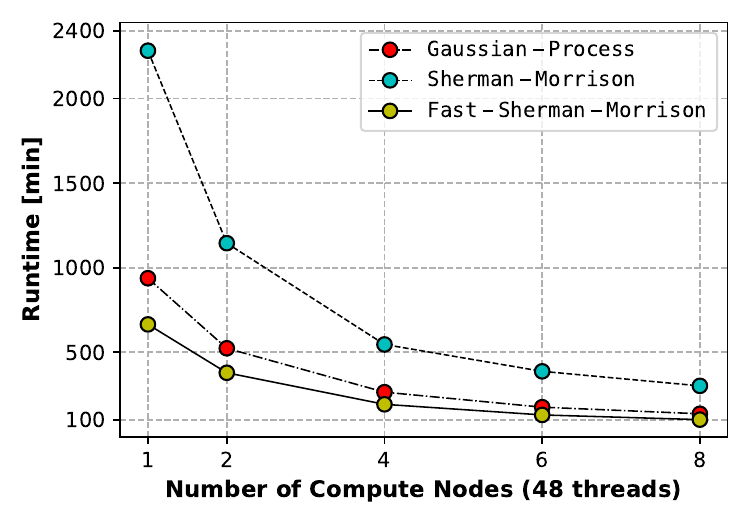}\label{fig:SPNA-ecorr-comp-2}}
\caption{Runtime comparison for \texttt{POCOMC} (left) and \texttt{PBILBY\_PTA} (right) methods for SPNA on the simulated dataset using the noise model described in Section \ref{subsec:Analysis-spna} with $N_c=100$ for all red noises and \texttt{ECORR} modeled using the GP (red), \texttt{sherman-morrison} (cyan) and \texttt{fast-sherman-morrison} (yellow) implementations.}
\label{fig:SPNA-ecorr-comp}
\end{figure}

\begin{figure}[!ht]
\centering
    \subfigure[Comparison of different sampling packages for SPNA]{\includegraphics[trim=0cm 0cm 0cm 0cm, clip, width=0.503\linewidth]{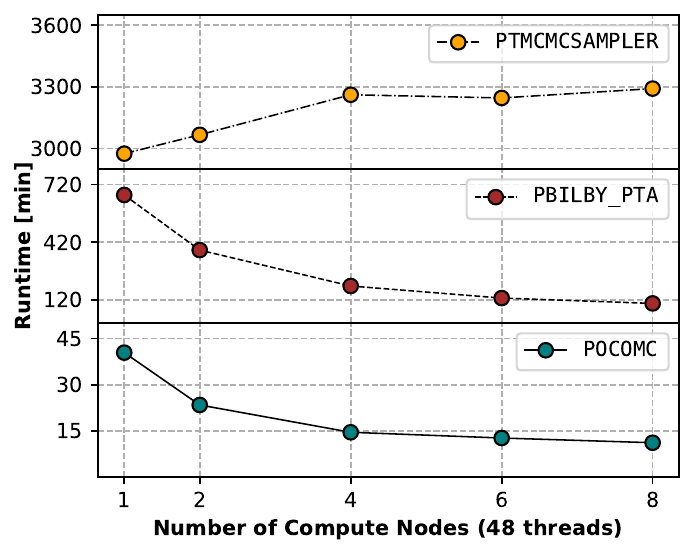}\label{fig:SPNA-model-comp}}
    \subfigure[\texttt{POCOMC} with adaptive $\rm n_{active}$]{\includegraphics[trim=0cm 0cm 0cm 0cm, clip, width=0.487\linewidth]{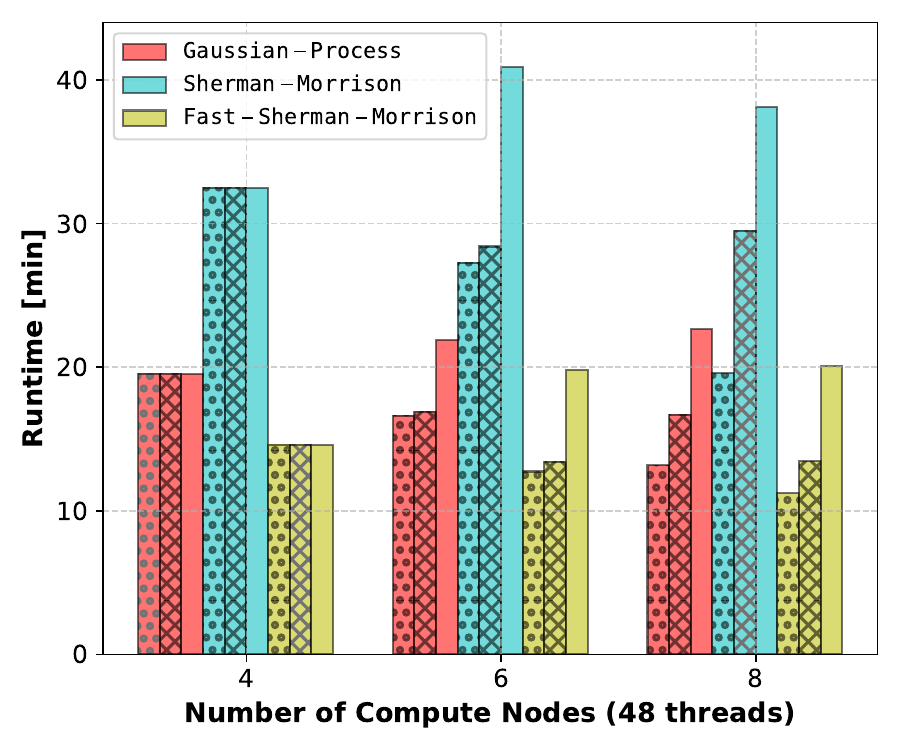}\label{fig:SPNA-comp-pocomc}}
\caption{\textit{Left}: Runtime comparison for SPNA with \texttt{fast-sherman-morisson} implementation of \texttt{ECORR} using \texttt{PTMCMCSAMPLER} (top panel), \texttt{PBILBY\_PTA} (middle panel), and \texttt{POCOMC} (bottom panel) samplers. \textit{Right}: Runtime comparison for SPNA with different \texttt{ECORR} implementations for $\rm n_{particles}=256$ (\textit{o}-hatched bars), $\rm n_{particles}\gtrsim N_{CPUs}$ (\textit{cross}-hatched bars), and $\rm n_{particles}\gg N_{CPUs}$ (solid bars) using \texttt{POCOMC} sampler. The red, cyan and yellow bars represent GP, \texttt{sherman-morrison}, and \texttt{fast-sherman-morrison} implementations of \texttt{ECORR}.}
\label{fig:SPNA-comp}
\end{figure}

We performed the analysis with increasing parallelisation, such that $\rm N_{CPUs}=48n$ where $n\in[1,8]$. The results from \texttt{POCOMC} are shown in Figure \ref{fig:SPNA-ecorr-comp-1} while those from \texttt{PBILBY\_PTA} are shown in Figure \ref{fig:SPNA-ecorr-comp-2}. We found similar trends in computation times for different \texttt{ECORR} models using either of the two methods. The \texttt{sherman-morrison} implementation is the most compute intensive, despite the extent of parallelisation involved, with GP being modest, and \texttt{fast-sherman-morrison} being the most efficient. We found that \texttt{PTMCMCSAMPLER} takes significantly longer computation times for SPNA, due to which we restricted it's comparison with \texttt{POCOMC} and \texttt{PBILBY\_PTA} for only a specific case where \texttt{fast-sherman-morrison} implementation of \texttt{ECORR} is used, since it is the most efficient of the three as per our previous results. The results of this comparative analysis are shown in Figure \ref{fig:SPNA-model-comp}, where it can be clearly seen that \texttt{POCOMC} is the most efficient, even for the analysis done on a single node, followed by \texttt{PBILBY\_PTA}, while \texttt{PTMCMCSAMPLER} is found to be not only the least efficient, but also to have an inverse scaling with increasing parallelisation. We examined the posterior distributions of noise parameters from different samplers, and they were found to be in close agreement. Further details of parameter comparisons are presented in Appendix \ref{sec:AppA}. \\

A close examination of Figure \ref{fig:SPNA-ecorr-comp-1} reveals a saturation in computation times for $n\gtrsim4$. However, as per \citet{Karamanis+2022a}, the parallelisation efficiency with \texttt{POCOMC} should scale linearly with $\rm N_{CPUs}$ as long as $\rm N_{CPUs}\leq n_{particles}$ (see Figure 3 therein). In the present work, for $n=6$, based on our compute architecture, we obtain $\rm N_{CPUs}=48\times6=288 > n_{particles}$. We reckon that this can be a plausible reason for the observed saturation. In order to investigate this further, we performed SPNA with all three \texttt{ECORR} implementations in two regimes -- $\rm n_{particles}=N_{CPUs}+\varepsilon$ for some $\varepsilon\ll\rm N_{CPUs}$\footnote{We have considered $\varepsilon=10$ without any loss of generality.} such that effectively $\rm n_{particles}\gtrsim N_{CPUs}$, and $\rm n_{particles}=2N_{CPUs}$ such that $\rm n_{particles}\gg N_{CPUs}$. The obtained results are shown in Figure \ref{fig:SPNA-comp-pocomc} where the \textit{o}-hatched bars represent the $\rm n_{particles}=256$ case, the \textit{cross}-hatched bars represent the $\rm n_{particles}\gtrsim N_{CPUs}$ case, and the solid bars represent the $\rm n_{particles}\gg N_{CPUs}$ case. We can see that all cases match for $n=4$ since that corresponds to $\rm N_{CPUs}=192 < n_{particles}$. However, for $n\geq6$, we find a systematic increase in computation times such that $\rm T_{(n_{particles}=256)} < T_{(n_{particles}\gtrsim N_{CPUs})} < T_{(n_{particles}\gg N_{CPUs})}$, with the results being agnostic to the \texttt{ECORR} implementation. We examine the potential causes for such results in Section \ref{sec:Discussion}. A similar transcendental saturation is also observed for \texttt{PBILBY\_PTA} in Figure \ref{fig:SPNA-ecorr-comp-2}, but that's not surprising since Nested Sampling algorithms are inherently sub-linear in scaling \citep{Handley+2015}.

\subsection{Common Red Noise Analysis}\label{subsec:Results-crn}

\begin{figure}[!ht]
\centering
    \subfigure[CURN analysis]{\includegraphics[trim=0cm 0cm 0cm 0cm, clip, width=0.495\linewidth]{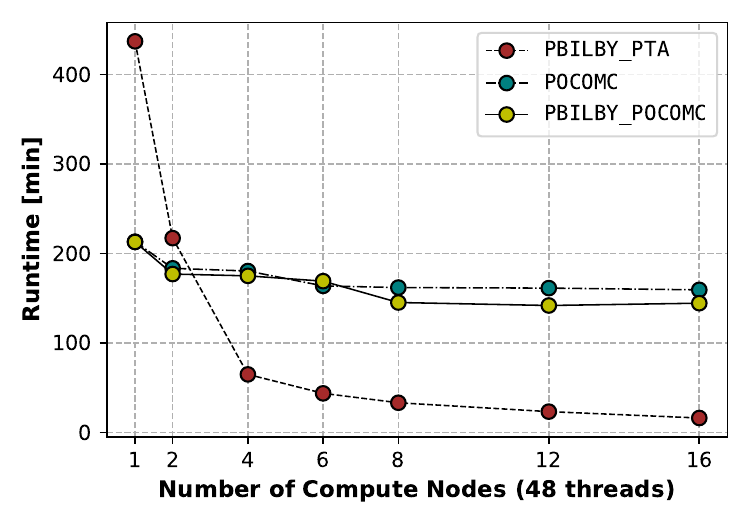}\label{fig:CRN-comp-1}}
    \subfigure[HD analysis]{\includegraphics[trim=0cm 0cm 0cm 0cm, clip, width=0.495\linewidth]{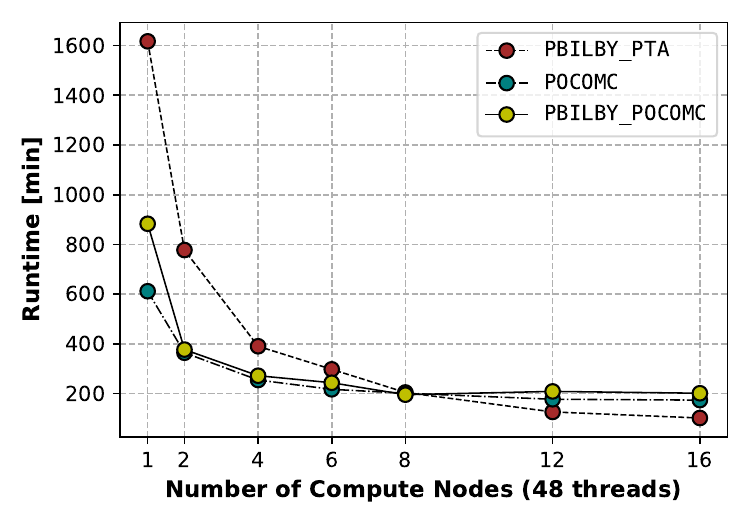}\label{fig:CRN-comp-2}}
\caption{Runtime comparison for CURN and HD analyses on a simulated dataset using \texttt{PBILBY\_PTA} (brown), \texttt{POCOMC} (teal) and \texttt{PBILBY\_POCOMC} (yellow) sampling methods. The SPNA framework and simulated dataset is as described in Section \ref{subsec:Analysis-crn}.}
\label{fig:CRN-comp}
\end{figure}

We performed the CRN analysis with both CURN and HD models using all three packages used for SPNA. However, we found that \texttt{PTMCMCSAMPLER} takes $\sim$$5\,\rm{d}$ for CURN search and $>$$10\,\rm{d}$ for HD search\footnote{The lower limit corresponds to the wall time on the cluster used for this work.} on a single compute node with 48 threads. Due to such large computation times (and sub-optimal scaling as per Figure \ref{fig:SPNA-model-comp}), we restrict ourselves to the comparison of \texttt{POCOMC} and \texttt{PBILBY\_PTA} only in this section. We have used $\rm n_{particles}=256$ for \texttt{POCOMC} as it was the most efficient setup according to the previous discussion, and the \texttt{PBILBY\_PTA} configuration is also set to the one described in Section \ref{subsec:Analysis-spna}. Additionally, we have compared these results with the \texttt{PBILBY\_POCOMC} implementation discussed in Section \ref{subsec:POCOMC} for completeness. Figure \ref{fig:CRN-comp-1} shows the runtime comparison for CURN search and Figure \ref{fig:CRN-comp-2} shows the same for HD search, done on the simulated dataset described in Section \ref{subsec:Analysis-crn} using \texttt{PBILBY\_PTA} (red), \texttt{POCOMC} (cyan) and \texttt{PBILBY\_POCOMC} (yellow) methods. We found that the parallelisation achieved using \texttt{PBILBY\_PTA} is significantly more efficient than that with \texttt{POCOMC}, with runtime varying from $\sim$$450\,\rm{min}$ to $\sim$$10\,\rm{min}$ for CURN and $\sim$$1600\,\rm{min}$ to $\sim$$100\,\rm{min}$ for HD, as the compute nodes increased from $n=1$ to $n=16$, respectively. \\

This is not observed for \texttt{POCOMC}, where the runtimes saturate at $\sim$$160\,\rm{min}$ for CURN and $\sim$$200\,\rm{min}$ for HD, with almost negligible parallelisation efficiency for the CURN analysis. Nevertheless, \texttt{POCOMC} performs significantly better than \texttt{PBILBY\_PTA} for $n\leq2$. The results with \texttt{PBILBY\_POCOMC} are found to be similar to those of \texttt{POCOMC}, except for the solitary case of HD analysis with $n=1$, where it was $\sim$$1.5$ times slower than \texttt{POCOMC}. We also used the adaptive $\rm n_{particles}$ scheme of Section \ref{subsec:Results-spna} and found similar scaling of runtime for both CURN and HD analyses. Further details of parameter comparisons are presented in Appendix \ref{sec:AppA}.

\section{Discussion}\label{sec:Discussion}
The results of SPNA and CRN analyses presented in the previous section show a remarkable improvement in computation time using \texttt{POCOMC} when it comes to single-node runs, which can be attributed to its novel NF-based preconditioning methodology with an ideal sampling efficiency as per the dimensionality of the problem \citep{Karamanis+2022a}. Compared to the widely adopted \texttt{PTMCMCSAMPLER}-based scheme, the decrease in runtime from $>$$10\,\rm{d}$ to $\sim$$10\,\rm h$ with a modest number of CPUs opens up the possibility of exploring more complicated models for GW analyses, which were earlier limited due to the computational bottleneck. Furthermore, the decrease in runtime from $\sim$$50\,\rm h$ with \texttt{PTMCMCSAMPLER} to $\sim$$40\,\rm m$ with \texttt{POCOMC} on a single node, along with the fact that a single run provides both MCMC chains as well as evidence estimates, makes it a suitable choice for future SPNA in PTA datasets. \\

However, the results presented for both SPNA and CRN analysis show deviation from the \textit{exact} linear scaling behavior discussed in \citet{Karamanis+2022a} for \texttt{POCOMC}. This \textit{exact} scaling predominantly arises from the \textit{exactly parallelisable} mutation step of PMC. In particular, the mutation step in \texttt{POCOMC} involves evolving a set of particles to geometrically interpolate between the prior and posterior densities via the likelihood, while employing an adaptive temperature scheduling scheme, parameterised by $\beta_{\rm T}$, such that the posterior is approximated as $\Pi\mathcal{L}^{\beta_{\rm T}}$. We suspect that the increase in computation time with $\rm n_{particles}$ for $n\geq 6$, as shown in Figure \ref{fig:SPNA-comp-pocomc}, could be due to \textit{overcrowding} of this mutation step. This means that the sequence of 256 particles is enough to adequately interpolate the posterior density. A further increase of $\rm n_{particles}$ with $\rm N_{CPUs}$ will lead to finer interpolation but will enhance the computation time, as more mutations are required to be computed at each temperature. Therefore, the observed saturation in runtimes seems to be a potential caveat of this trade-off, and $\rm n_{particles}=256$ turns out to be the optimal choice. This is further substantiated by the results in Appendix \ref{sec:AppB}, wherein no improvement in parameter constraints is observed from the posteriors with increased $\rm n_{particles}$. \\

On the other hand, \texttt{PBILBY\_PTA} achieves the most efficient parallelisation up to the extent tested in this study, with an exponential improvement in runtime for all the presented analyses. This facilitated a remarkable decrease in runtime for some of the most intensive computations in pulsar timing, even outperforming \texttt{POCOMC} when it comes to massively-parallelised CRN analyses. Therefore, this architecture seems to be the optimal choice for highly compute-intensive GW analyses on HPC facilities with massive parallelisation capabilities, which are now becoming ubiquitous in the modern world. These significant accelerations not only facilitate faster scientific deliverables, but also encourage the test of more complicated models to expand the scientific horizon. Based on the presented work, \texttt{POCOMC} serves as the optimal choice for single-node or SPNA computations, while \texttt{PBILBY\_PTA} is best suited for ensemble-level PTA analyses on architectures supporting massive parallelisation.

\section{Summary}\label{sec:Summary}
In this work, we presented a comparative study of computation times for intensive pulsar timing analyses, in particular, SPNA and CRN process searches in a fully Bayesian framework. In order to estimate the posterior distributions of parameters, we used two of the most favored packages in pulsar timing community, namely \texttt{PTMCMCSAMPLER} \citep{Ellis+2017} for MCMC analysis, and \texttt{DYNESTY} \citep{Skilling2004} for Nested Sampling. For the first time, in this work, we have applied the \texttt{POCOMC} \citep{Karamanis+2022a, Karamanis+2022b} package to PTA studies. The uniqueness of \texttt{POCOMC} is its Normalising Flow-based preconditioning, which has the potential to significantly reduce the computational cost, especially for problems involving high-dimensionality and strong correlations -- a regime where most PTA analyses suffer from insatiable computational bottlenecks at present. We incorporated Nested Sampling with the parallelisation architecture of \texttt{PARALLEL\_BILBY} \citep{Smith+2020, Samajdar+2022}, referred to as \texttt{PBILBY\_PTA} in this work. We also extended the same to include \texttt{POCOMC} within the parallelisation architecture of \texttt{PARALLEL\_BILBY}, referred to as \texttt{PBILBY\_POCOMC} in this work. We used the \texttt{ENTERPRISE} \citep{Ellis+2020, Johnson+2024} framework to model the PTA likelihood on simulated datasets curated for SPNA and CRN analysis, where both CURN and HD-correlated searches are included. The SPNA dataset included the spin, DM and SW red noise processes along with \texttt{EFAC}, \texttt{EQUAD} and \texttt{ECORR} for white noise, where we compared the SPNA recovery using the GP \citep{Lentati+2014}, \texttt{sherman-morrison} \citep{ShermanMorrison1949, ShermanMorrison1950} and \texttt{fast-sherman-morrison} \citep{vanHaasteren2023} implementations of \texttt{ECORR}. The CRN dataset was simulated as per \citep{Pol+2021} using the characteristics of the InPTA DR2 dataset \citep{Rana+2025, Nobleson+2026}, with additional GWB injections. For all the analyses, we compared the parallelisation capability and its efficiency achieved by all the sampling methods by distributing the computations across multiple compute nodes of an HPC cluster with MPI-enabled communication. \\

We found that amongst the three \texttt{ECORR} implementations, \texttt{fast-sherman-morrison} is the most efficient and \texttt{sherman-morrison} is the least, across all the sampling techniques. Furthermore, we found \texttt{PTMCMCSAMPLER} to be the most expensive in terms of runtime, followed by \texttt{PBILBY\_PTA}, and \texttt{POCOMC} being the fastest. However, the parallelisation efficiency was highest for \texttt{PBILBY\_PTA}, where the runtime reduces by a factor of 6 as the processes were distributed across 8 compute nodes, compared to a factor of 3 reduction with \texttt{POCOMC}. That said, the single node runtime with \texttt{POCOMC} outperformed 8 node runtime of \texttt{PBILBY\_PTA} by a factor of 3. We compared only these two and \texttt{PBILBY\_POCOMC} for more intensive CRN searches, since \texttt{PTMCMCSAMPLER} takes several orders longer computation times. For CRN searches also, we found the most efficient parallelisation being achieved by \texttt{PBILBY\_PTA}, however, \texttt{POCOMC} still being the fastest in single node performance. \texttt{PBILBY\_POCOMC} was found to scale similar to \texttt{POCOMC}. We also found a saturation of runtime with parallelisation beyond 4 nodes in \texttt{POCOMC} with a fixed set of 256 particles, and an increased runtime for the set increasing with the number of processors. We suspect that this could be due an effective trade-off at the overcrowded mutation step of preconditioning. Nevertheless, our comparative study suggests \texttt{POCOMC} to be a significant improvement over the existing sampling techniques used in the pulsar timing community, especially when it comes to analysis done without any HPC facility. The use of \texttt{PBILBY\_PTA} is most effective when HPC architectures with a large set of communicating nodes are employed.

\section*{Acknowledgements}\label{sec:Acknowledgements}
The work of CD at the Physical Research Laboratory (PRL) was supported by the Department of Space, Government of India. CD acknowledges the Param Vikram-1000 High Performance Computing Cluster of the Physical Research Laboratory (PRL), a unit of the Department of Space, Government of India, for performing the intensive computations. CD acknowledges RM for the fruitful discussions and support during the work. HS acknowledges the Stepwell High-Performance Computing Facility at Ahmedabad University for providing the requisite computational resources. HT is supported by DST INSPIRE fellowship.

\section*{Software}\label{sec:Software}
\begin{itemize*}
    \item \texttt{PYTHON} \citep{vanRossum+2009}
    \item \texttt{ASTROPY} \citep{Whelan+2022}
    \item \texttt{SCIPY} \citep{Virtanen+2020}
    \item \texttt{MATPLOTLIB} \citep{Hunter2007}
    \item \texttt{CORNER} \citep{Foreman-Mackey2016}
    \item \texttt{NUMPY} \citep{Harris+2020}
    \item \texttt{TEMPO2} \citep{Hobbs+2006, Edwards+2006}
    \item \texttt{PINT} \citep{Luo+2021, Susobhanan+2024}
    \item \texttt{ENTERPRISE} \citep{Ellis+2020, Johnson+2024}
    \item \texttt{ENTERPRISE\_EXTENSIONS} \citep{Taylor+2021}
    \item \texttt{PTMCMCSAMPLER} \citep{Ellis+2017}
    \item \texttt{DYNESTY} \citep{Speagle2020}
    \item \texttt{PARALLEL\_BILBY} \citep{Smith+2020}
    \item \texttt{PBILBY\_PTA} \citep{Samajdar+2022}
    \item \texttt{POCOMC} \citep{Karamanis+2022a, Karamanis+2022b}
\end{itemize*}

\section{Data Availability}\label{sec:Data}
The entire work presented in this paper is based on simulated datasets, generated solely for this purpose using publicly available software. The SPNA dataset is shared along with the manuscript as supplementary material\update{\dataset[here]{https://doi.org/10.5281/zenodo.23058320}}.

\bibliography{parallel}
\bibliographystyle{aasjournalv7}

\appendix

\section{Benchmarking noise recovery of \texttt{POCOMC} with \texttt{PTMCMCSAMPLER} and \texttt{PBILBY\_PTA}}\label{sec:AppA}
\setcounter{figure}{0}
\renewcommand{\thefigure}{A\arabic{figure}}

\begin{figure}[!ht]
    \centering
    \includegraphics[trim=0cm 0cm 0cm 0cm, clip, width=\linewidth]{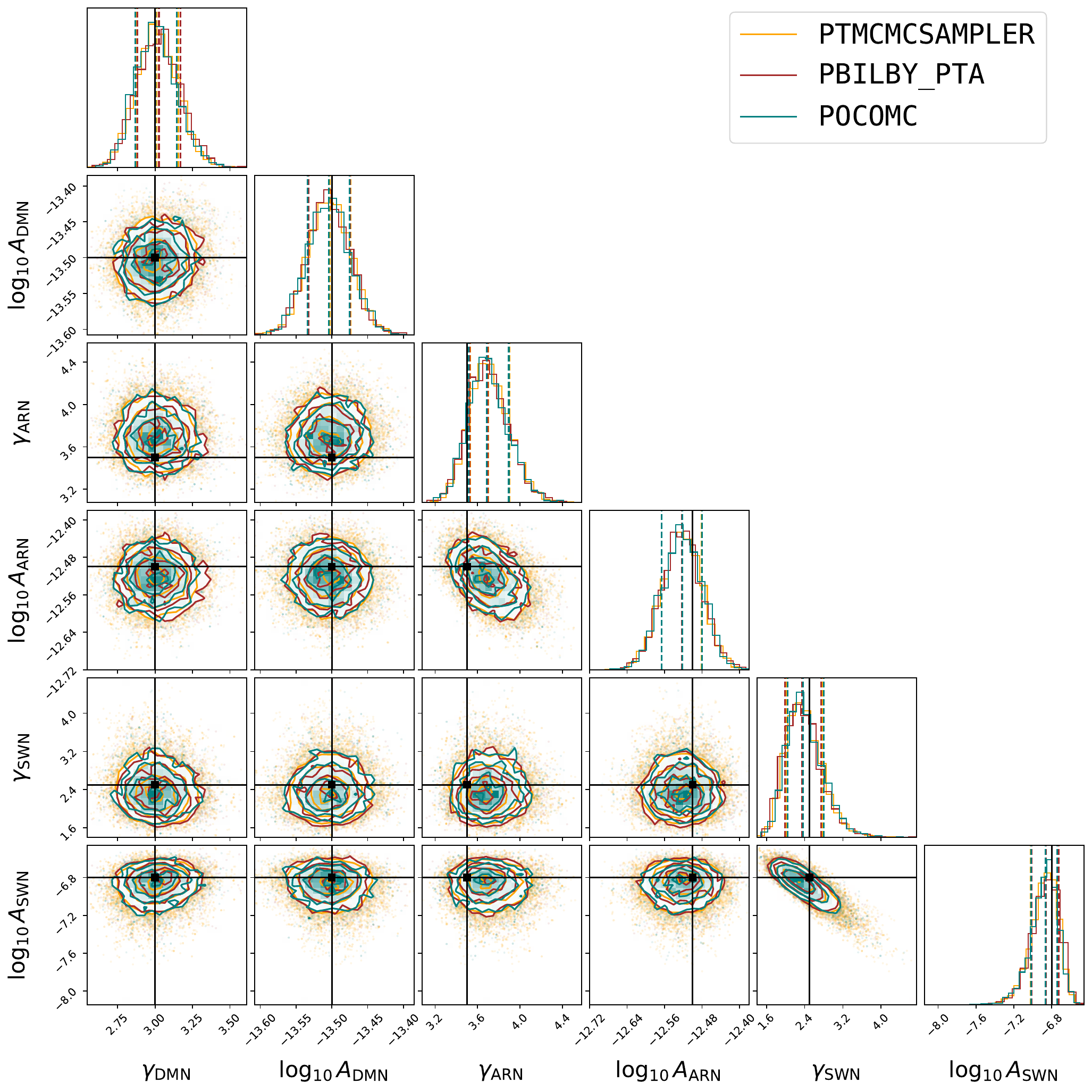}
    \caption{Posterior distributions of single-pulsar red noise hyper-parameters obtained using the \texttt{PTMCMSAMPLER} (orange), \texttt{PBILBY\_PTA} (brown), and \texttt{POCOMC} (teal) packages, with \texttt{fast-sherman-morrison} implementation of \texttt{ECORR} in SPNA. The injected single-pulsar noise processes are as described in Section \ref{subsec:Analysis-spna} and are highlighted as black solid lines in the respective posterior panels.}
    \label{fig:SPNA-posteriors}
\end{figure}

Figure \ref{fig:SPNA-posteriors} shows the posterior distributions of single-pulsar red noise hyper-parameters obtained with the SPNA analysis on the simulated dataset discussed in Section \ref{subsec:Analysis-spna} using the \texttt{PTMCMSAMPLER} (orange), \texttt{PBILBY\_PTA} (brown), and \texttt{POCOMC} (teal) packages. Only a representative case with \texttt{fast-sherman-morrison} implementation of \texttt{ECORR} in SPNA for $n=1$ is shown for all the packages.

\begin{figure}[!ht]
    \centering
    \subfigure[CURN analysis]{\includegraphics[trim=0cm 0cm 0cm 0cm, clip, width=0.493\linewidth]{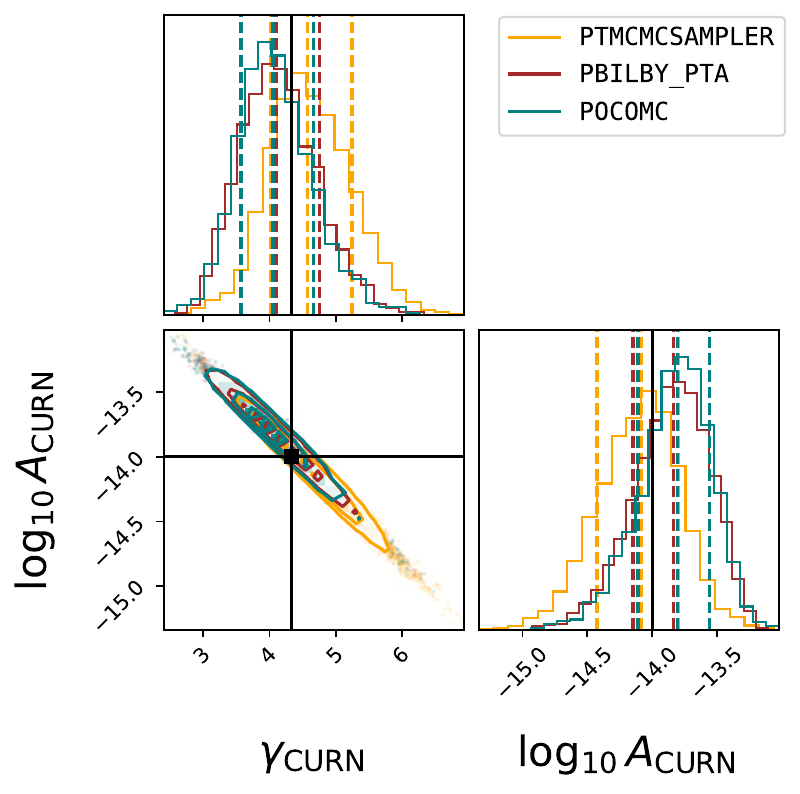}\label{fig:CURN-posteriors}}
    \subfigure[HD analysis]{\includegraphics[trim=0cm 0cm 0cm 0cm, clip, width=0.497\linewidth]{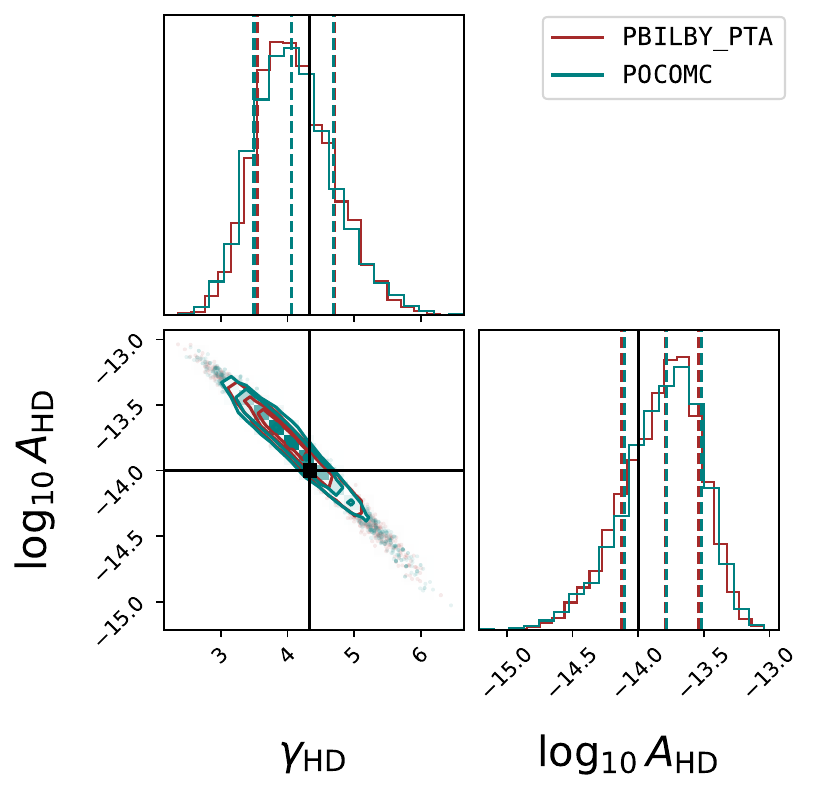}\label{fig:HD-posteriors}}
\caption{Posterior distributions of CURN process (left) and HD-correlated process (right) spectral hyper-parameters obtained using the \texttt{PTMCMSAMPLER} (orange), \texttt{PBILBY\_PTA} (brown), and \texttt{POCOMC} (teal) packages, for the CRN analysis. The injected single-pulsar noise processes and simulated dataset are as described in Section \ref{subsec:Analysis-crn} and are highlighted as black solid lines in the respective posterior panels. Only results for CURN analysis with \texttt{PTMCMCSAMPLER} are shown.}
\label{fig:CRN-posteriors}
\end{figure}

Figure \ref{fig:CRN-posteriors} shows the posterior distributions of CURN (left) and HD-correlated (right) process spectral hyper-parameters obtained with the CRN analysis on the simulated dataset discussed in Section \ref{subsec:Analysis-crn} using the \texttt{PTMCMSAMPLER} (orange), \texttt{PBILBY\_PTA} (brown), and \texttt{POCOMC} (teal) packages. Only a representative case with $n=4$ is shown for the \texttt{PBILBY\_PTA} and \texttt{POCOMC} packages, while the \texttt{PTMCMSAMPLER} run was done with $n=1$. Since the runtime with \texttt{PTMCMCSAMPLER} for HD analysis was beyond the permissible wall time ($>10\,\rm d$) of the cluster used in this work, no posteriors for the same are plotted in Figure \ref{fig:HD-posteriors}. While the \texttt{PTMCMCSAMPLER} posteriors for CURN process seem to be in slight tension with the rest, those from \texttt{PBILBY\_PTA} and \texttt{POCOMC} seem to be in good agreement.

\newpage

\section{Posteriors for varying particle sample in \texttt{POCOMC}}\label{sec:AppB}
\setcounter{figure}{0}
\renewcommand{\thefigure}{B\arabic{figure}}

\begin{figure}[!ht]
    \centering
    \includegraphics[trim=0cm 0cm 0cm 0cm, clip, width=\linewidth]{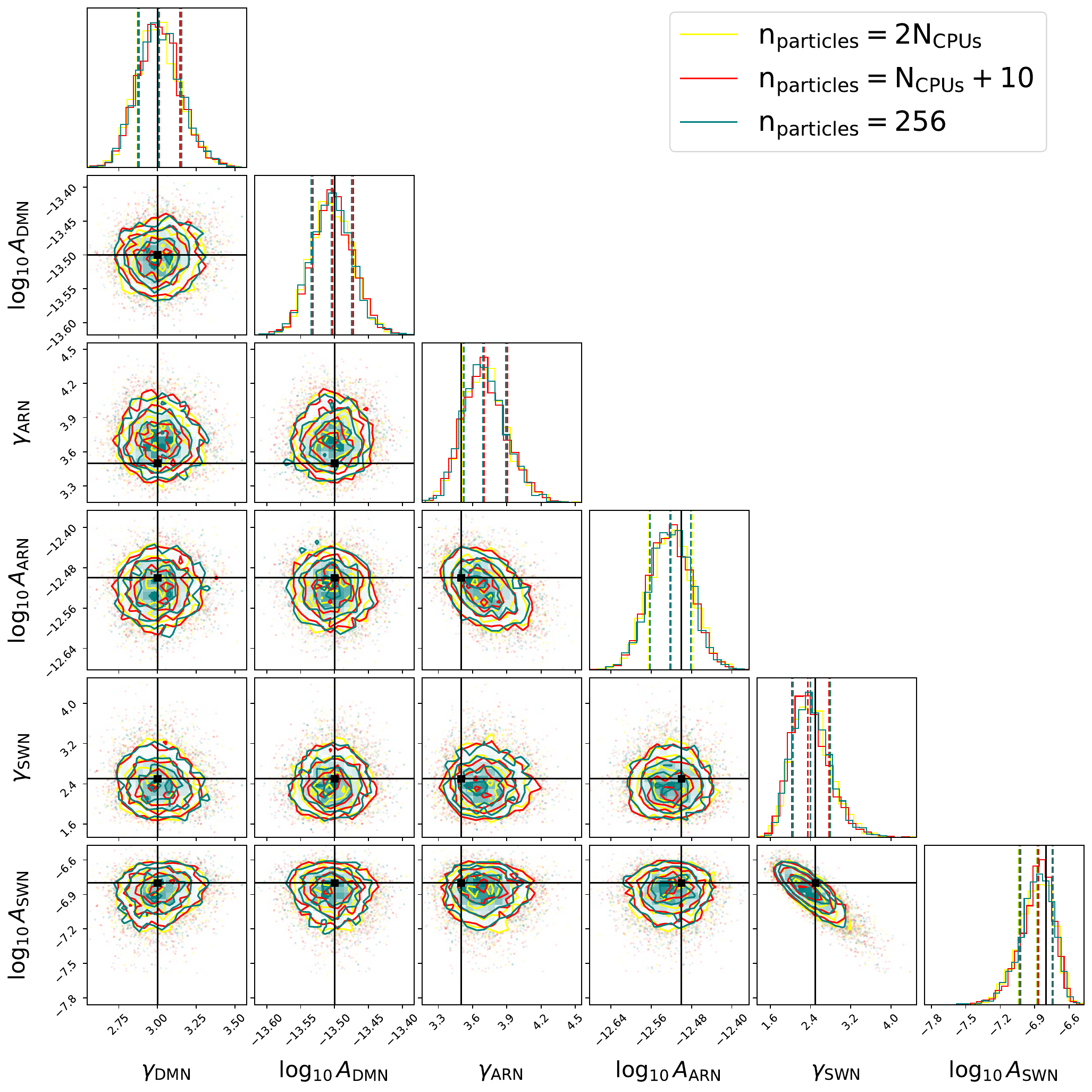}
    \caption{Posterior distributions of single-pulsar red noise hyper-parameters obtained using the \texttt{POCOMC} package for $\rm n_{particles}=2N_{CPUs}$ (yellow), $\rm n_{particles}=N_{CPUs}+10$ (red) and $\rm n_{particles}=256$ (teal), with \texttt{fast-sherman-morrison} implementation of \texttt{ECORR} in SPNA. The injected single-pulsar noise processes are as described in Section \ref{subsec:Analysis-spna} and are highlighted as black solid lines in the respective posterior panels.}
    \label{fig:SPNA-varyingN-posteriors}
\end{figure}

The SPNA posteriors obtained using the \texttt{POCOMC} package for $\rm n_{particles}=2N_{CPUs}$ (yellow), $\rm n_{particles}=N_{CPUs}+10$ (red) and $\rm n_{particles}=256$ (teal), with \texttt{fast-sherman-morrison} implementation of \texttt{ECORR} (refer Section \ref{subsec:Analysis-spna}) are shown in Figure \ref{fig:SPNA-varyingN-posteriors}, for a representative case of $n=8$. No statistically significant differences can be seen in the estimated posteriors for very large $\rm n_{particles}$. Therefore, the choice of $\rm n_{particles}=256$ seems to be the most optimal.

\end{document}